\documentclass[11pt]{article}\textwidth 6.5in\textheight 9in
\usepackage{amssymb}\usepackage[colorlinks]{hyperref}\usepackage{color}
	\usepackage{stmaryrd}\usepackage{mathrsfs}
	\usepackage{graphicx}
	\usepackage{amsmath}
\usepackage{caption}

\begin{document}
\setlength{\captionmargin}{27pt}
\newcommand\hreff[1]{\href {http://#1} {\small http://#1}}
\newcommand\trm[1]{{\bf\em #1}} \newcommand\emm[1]{{\ensuremath{#1}}}
\newcommand\prf{\paragraph{Proof.}}\newcommand\qed{\hfill\emm\blacksquare}

\newtheorem{thr}{Theorem} 
\newtheorem{lmm}{Lemma}
\newtheorem{cor}{Corollary}
\newtheorem{con}{Conjecture} 
\newtheorem{prp}{Proposition}

\newtheorem{blk}{Block}
\newtheorem{dff}{Definition}
\newtheorem{asm}{Assumption}
\newtheorem{rmk}{Remark}
\newtheorem{clm}{Claim}
\newtheorem{exm}{Example}

\newcommand\Ks{\mathbf{Ks}} 
\newcommand{\ab}{a\!b}
\newcommand{\yx}{y\!x}
\newcommand{\yux}{y\!\underline{x}}

\newcommand\floor[1]{{\lfloor#1\rfloor}}\newcommand\ceil[1]{{\lceil#1\rceil}}

\newcommand{\lea}{<^+}
\newcommand{\gea}{>^+}
\newcommand{\eqa}{=^+}

\newcommand{\lel}{<^{\log}}
\newcommand{\gel}{>^{\log}}
\newcommand{\eql}{=^{\log}}

\newcommand{\lem}{\stackrel{\ast}{<}}
\newcommand{\gem}{\stackrel{\ast}{>}}
\newcommand{\eqm}{\stackrel{\ast}{=}}

\newcommand\edf{{\,\stackrel{\mbox{\tiny def}}=\,}}
\newcommand\edl{{\,\stackrel{\mbox{\tiny def}}\leq\,}}
\newcommand\then{\Rightarrow}

\newcommand\C{\mathbf{C}} 

\renewcommand\chi{\mathcal{H}}
\newcommand\km{{\mathbf {km}}}\renewcommand\t{{\mathbf {t}}}
\newcommand\KM{{\mathbf {KM}}}\newcommand\m{{\mathbf {m}}}
\newcommand\md{{\mathbf {m}_{\mathbf{d}}}}\newcommand\mT{{\mathbf {m}_{\mathbf{T}}}}
\newcommand\K{{\mathbf K}} \newcommand\I{{\mathbf I}}

\newcommand\II{\hat{\mathbf I}}
\newcommand\Kd{{\mathbf{Kd}}} \newcommand\KT{{\mathbf{KT}}} 
\renewcommand\d{{\mathbf d}} 
\newcommand\D{{\mathbf D}}

\newcommand\w{{\mathbf w}}
\newcommand\Cs{\mathbf{Cs}} \newcommand\q{{\mathbf q}}
\newcommand\E{{\mathbf E}} \newcommand\St{{\mathbf S}}
\newcommand\M{{\mathbf M}}\newcommand\Q{{\mathbf Q}}
\newcommand\ch{{\mathcal H}} \renewcommand\l{\tau}
\newcommand\tb{{\mathbf t}} \renewcommand\L{{\mathbf L}}
\newcommand\bb{{\mathbf {bb}}}\newcommand\Km{{\mathbf {Km}}}
\renewcommand\q{{\mathbf q}}\newcommand\J{{\mathbf J}}
\newcommand\z{\mathbf{z}}

\newcommand\B{\mathbf{bb}}\newcommand\f{\mathbf{f}}
\newcommand\hd{\mathbf{0'}} \newcommand\T{{\mathbf T}}
\newcommand\R{\mathbb{R}}\renewcommand\Q{\mathbb{Q}}
\newcommand\N{\mathbb{N}}\newcommand\BT{\{0,1\}}
\newcommand\FS{\BT^*}\newcommand\IS{\BT^\infty}
\newcommand\FIS{\BT^{*\infty}}
\renewcommand\S{\mathcal{C}}\newcommand\ST{\mathcal{S}}
\newcommand\UM{\nu_0}\newcommand\EN{\mathcal{W}}

\newcommand{\supp}{\mathrm{Supp}}

\newcommand\lenum{\lbrack\!\lbrack}
\newcommand\renum{\rbrack\!\rbrack}

\newcommand\h{\mathbf{h}}
\renewcommand\qed{\hfill\emm\square}
\renewcommand\i{\mathbf{i}}
\newcommand\p{\mathbf{p}}
\renewcommand\q{\mathbf{q}}
\title{\vspace*{-3pc} The Outlier Theorem Revisited}

\author {Samuel Epstein\footnote{JP Theory Group. samepst@jptheorygroup.org}}

\maketitle

\begin{abstract}
	An outlier is a datapoint that is set apart from a sample population.  The outlier theorem in algorithmic information theory states that given a computable sampling method, outliers must appear. We present a simple proof to the outlier theorem, with exponentially improved bounds. We extend the outlier theorem to ergodic dynamical systems which are guaranteed to hit ever larger outlier states with diminishing measures.  We show how to construct deterministic functions from random ones, i.e. function derandomization. We also prove that all open sets of the Cantor space with large uniform measure will either have a simple computable member or high mutual information with the halting sequence.
\end{abstract}
\section{Introduction}
The deficiency of randomness of an infinite sequence $\alpha\in\IS$ with respect to a computable measure $P$ over $\IS$ is defined to be $\D(\alpha|P)=\sup_n-\log P(\alpha[0..n])-\K(\alpha[0..n])$. The term $\K$ is the prefix free Kolmogorov complexity. \\

\noindent\textbf{Theorem A.} \textit{For computable measures $\mu$ and non-atomic $\lambda$ over $\IS$ and $n\in\N$,\\ $\lambda\{\alpha :\D(\alpha|\mu)>n\}> 2^{-n-\K(n,\mu,\lambda)-O(1)}$. }\\

This equation has special meaning when $\lambda$ is the stationary measure of a dynamical system. The theorem was proven using a general template consistent with the Independence Postulate, \cite{Levin13,Levin84}. This method involves first proving that an object has mutual information with the halting sequence. The second step involves removing the mutual information term from the inequality. The removal of the information term can be done in a number of ways, and the dynamical systems theorem represents one such example.

\subsection{Outliers} In addition, we present a simple proof of the outlier theorem in \cite{Epstein21} with exponentially improved bounds. A sampling method $A$ is a probabilistic function that maps an integer $N$ with probability 1 to a set containing $N$ different strings. Let $P=P_1,P_2,\dots$ be a sequence of measures over strings. For example, one may choose $P_1=P_2\dots$ or choose $P_n$ to be the uniform measure over $n$-bit strings. A conditional probability bounded $P$-test is a function $t:\FS\times\N\rightarrow \R_{\geq 0}$ such that for all $n\in\N$ and positive real number $r$, we have $P_n(\{x:t(x|n)\geq r\})\leq 1/r$. If $P_1, P_2,\dots$ is uniformly computable, then there exists a lower-semicomputable such $P$-test $t$ that is ``maximal'' (i.e., for which $t'\leq O(t)$ for every other such test  $t'$). We fix such a $t$, and let $\overline{\d}_n(x|P) = \log t(x|n)$.\\

\noindent\textbf{Theorem B.} \textit{Let $P=P_1,P_2\dots$ be a uniformly computable sequence of measures on strings and let $A$ be a sampling method. There exists $c\in\N$ such that for all $n$ and $k$:
	$$
	\Pr\left(\max_{a\in A(2^n)}\overline{\d}_n(a|P)> n - k -c\right)\geq 1-2e^{-2^k}.
	$$}

\subsection{Function Derandomization}
 In this paper, we show how to construct deterministic functions from random ones. Random functions $F$ over natural numbers are modeled by discrete stochastic processes indexed by $\N$, where each $F(t)$, $t\in\N$, is a random variable over $\N$. $\mathcal{F}$ is the set of all random functions. A 
random function $F\in\mathcal{F}$ is computable if there is a program that on input $(a_1,\dots,a_n)$ lower computes $\Pr\left[F(1)=a_1\cap F(2)=a_2\cap\dots \cap F(n)=a_n \right]$. Put another way, a random function $F\in\mathcal{F}$ is computable if $X=\Pr[F(a_1)=b_1 \cap\dots\cap F(a_n)=b_n]$ is uniformly computble in $\{(a_i,b_i)\}_{i=1}^n$.  
 The complexity $\K(F)$ of a random function $F\in\mathcal{F}$, is the smallest program that
computes $X$. $\mathcal{G}$ is the set of all deterministic functions $G:\N\rightarrow\N$. A sample $S\in\mathcal{S}$ is a finite set of pairs $\{(a_i,b_i)\}_{i=1}^n$. $\mathcal{S}$ is the set of all samples.  The encoding of a sample is $\langle S\rangle= \langle \{(a_i,b_i)\}_{i=1}^n\rangle$.  We say $G(S)$ if $G$ is consistent with $S$, with $G(a_i)=b_i$, $i=1,\dots,n$. For random functions, $F(S)$ is the event that $F$ is consistent with $S$. The amount of information that a string has with the halting sequence $\ch\in\IS$ is $\I(x;\ch)=\K(x)-\K(x|\ch)$.\\
 
\noindent\textbf{Theorem C.}\textit{
 For $F\in\mathcal{F}$, $S\in\mathcal{S}$,
 $\min_{G\in\mathcal{G},G(S)}\K(G) \lel \K(F) -\log \Pr[F(S)] + \I(\langle S\rangle;\mathcal{H})$.}
 
\subsection{Open Sets}  For $x\in\FS$ let $\Gamma_x=\{x\beta:\beta\in\IS\}$ be the interval of $x$. For open set $S\subseteq\IS$, let its encoding be $\langle S\rangle=\langle\{x:\Gamma_x\textrm{ is maximal in }S\}\rangle$.  Arbitrary open sets $S\subseteq\IS$ can have infinite $\langle S\rangle$. The Kolmogorov complexity of an infinite sequence $\alpha\in\IS$ is $\K(\alpha)$, the size of the smallest program to a universal Turing machine that will output, without halting, $\alpha$ on the output tape. Let $\mu$ be the uniform measure of the Cantor space. The information term between infinite sequences is $\I(\alpha:\beta)=\log\sum_{x,y\in\FS}\m(x|\alpha)\m(y|\beta)2^{\I(x:y)}$, where $\m$ is the algorithmic probability \cite{Levin74}. The mutual information between two finite strings is $\I(x:y)=\K(x)+\K(y)-\K(x,y)$.\\

\noindent\textbf{Theorem D.} \textit{For open $S\subseteq\IS$, $\min_{\alpha\in S}\K(\alpha)\lel -\log\mu(S)+\I(\langle S\rangle:\mathcal{H})$.}

\subsection{Other Results}

Theorems C and D are variations of the main theorem in \cite{Levin16,Epstein19}, but they are not directly implied by it. We discuss sampling methods over infinite sequences as well as non-halting sampling methods. We prove slightly stronger results to Theorem D for clopen sets. Derandomization can be generalized to sets of samples, and also to lower computable random functions. We apply function derandomization to games, showing how to create deterministic agents from probabilistic ones, with application to graph navigation. We show how derandomization can used to compress approximate solutions to NP hard problems, in particular {\sc Max-Cut} and {\sc Max-3Sat}. A monotone complexity variant to the main theorem in \cite{Levin16,Epstein19} is proven.  We also show that there is no equivalent to Theorem D for closed sets. Due to Anonymous, there exists closed sets $C\subset\IS$ with no computable members, $\mu(C)>0$, and $\I(\langle C\rangle:\mathcal{H})<\infty$.
\section{Conventions}
Let $\N$, $\Q$, $\R$, $\BT$, $\FS$, and $\IS$ be the sets of natural numbers, rationals, real numbers, bits, finite strings, and infinite strings. We use $\langle x\rangle$ to represent a self-delimiting code for $x\in\FS$, such as $1^{\|x\|}0x$. The self-delimiting code for a finite set of strings $\{a_i\}_{i=1}^n$ is $\langle\{a_i\}_{i=1}^n\rangle=\langle n\rangle\langle a_1\rangle\langle a_2\rangle\dots\langle a_n\rangle$. For $x\in\FS\cup\IS$ and $y\in\FS\cup\IS$, we use $x\sqsubseteq y$ if there is some string $z\in\FS\cup\IS$ where $xz=y$. We say $x\sqsubset y$ if $x\sqsubseteq y$ and $x\neq y$. For a mathematical statement $W$, $[W]=1$ if $W$ is true, and $[W]=0$ otherwise.

For positive real functions $f$ the terms  ${\lea}f$, ${\gea}f$, ${\eqa}f$ represent ${<}f{+}O(1)$, ${>}f{-}O(1)$, and ${=}f{\pm}O(1)$, respectively. In addition ${\lem}f$, ${\gem}f$ denote $<f/O(1)$, $>f/O(1)$. The term ${\eqm}f$  denotes ${\lem}f$ and ${\gem}f$. For the nonnegative real function $f$, the terms ${\lel}f$, ${\gel} f$, and ${\eql}f$ represent the terms ${<}f{+}O(\log(f{+}1))$, ${>}f{-}O(\log(f{+}1))$, and ${=}f{\pm}O(\log(f{+}1))$, respectively. 

A semi measure is a function $Q:\N\rightarrow\R_{\geq 0}$ such that $\sum_{a\in\N} Q(a)\leq 1$. A probability measure is a semi measure such that $\sum_{a\in\N} Q(a)= 1$. A probability measure $Q$ is elementary if $|\{a:Q(a)>0\}|<\infty$ and $\mathrm{Range}(Q)\subseteq\Q_{\geq 0}$. Elementary measures $Q$ can be encoded into finite strings $\langle Q\rangle$.

The universal probability of a string $x\in\FS$, conditional on $y\in\FS\cup\IS$, is $\m(x|y)=\sum\{2^{-\|p\|}:U_y(p)=x\}$. The coding theorem states $-\log \m(x|y)\eqa\K(x|y)$. By the chain rule $\K(x,y)\eqa \K(x)+\K(y|x,\K(x))$. The mutual information of a string $x$ with the halting sequence $\mathcal{H}\in\IS$ is $\I(x;\mathcal{H})=\K(x)-\K(x|\mathcal{H})$.

This paper uses notions of stochasticity in the field of algorithmic statistics \cite{VereshchaginSh17}. A string $x$ is stochastic, i.e. has a low $\Ks(x)$ score, if it is typical of a simple probability distribution. The deficiency of randomness function of a string $x$ with respect to an elementary probability measure $P$ conditional to $y\in\FS$, is $\d(x|P,y)=\floor{-\log P(x)}-\K(x|\langle P\rangle,y)$. 
\begin{dff}[Stochasticity]$ $
For $x,y\in\FS$,\\  $\Ks(x|y)=\min \{\K(P|y)+3\log \max\{\d(x|P,y),1\}: P\textrm { is an elementary probability measure}\}$. $\Ks(x)=\Ks(x|\emptyset)$. $\Ks(a|b) < \Ks(a) + O(\log \K(b))$.
\end{dff}

\section{Dynamical Systems}
In this section, we prove that dynamical systems will hit ever larger outliers with diminishing probability. To achieve this, we use the properties of the mutual information of an infinite sequence with the halting problem. The deficiency of randomness of an infinite sequence $\alpha\in\IS$ with respect to a computable probability measure $P$ over $\IS$ is defined to be $$\D(\alpha|P,x)=\log\sup_n\m(\alpha[0..n]|x)/P(\alpha[0..n]).$$ 
We have $\D(\alpha|P)=\D(\alpha|P,\emptyset)$. We require the following two theorems for the primary proof of this section.

\begin{thr}[\cite{Vereshchagin21,Levin74,Geiger12}]
	\label{thr:coninfo}
	$\Pr_{\mu}(\I(\alpha:\ch)>n)\lem 2^{-n+\K(\mu)}$.
\end{thr}

\begin{thr}[\cite{Epstein21}]
	\label{thr:InfExt} 
	For computable probability measure $P$ over $\IS$, for $Z\subseteq \IS$,  if $\N\ni s< \log\sum_{\alpha\in Z}2^{\D(\alpha|P)}$, then $s<\sup_{\alpha\in Z}\D(\alpha|P)\,{+}\,\I(\langle Z\rangle:\ch)+O(\K(s)+\log\I(\langle Z\rangle:\ch)+\K(P))$.
\end{thr}$ $\newpage

\begin{thr}[Dynamical Systems]
	\label{thr:dyn}
	For computable measures $\mu$ and nonatomic $\lambda$ over $\IS$ and $n\in\N$, $\lambda\{\alpha :\D(\alpha|\mu)>n\}> 2^{-n-\K(n,\mu,\lambda)-O(1)}$. 
\end{thr}
\begin{prf}
	We first assume not. For all $c\in\N$, there exist computable nonatomic measures $\mu$, $\lambda$, and there exists $n$,  where $\lambda\{\alpha :\D(\alpha|\mu)>n\}\leq 2^{-n-\K(n,\mu,\lambda)-c}$. Sample $2^{n+\K(n,\mu,\lambda)+c-1}$ elements $D\subset\IS$ according to $\lambda$. The probability that all samples $\beta\in D$ have $\D(\beta|\mu)\leq n$ is
	$$\prod_{\beta\in D}\lambda\{\D(\beta|\mu)\leq n\}\geq (1-|D|2^{-n-\K(n,\mu,\lambda)-c})\geq (1-2^{n+\K(n,\mu,\lambda)+c-1}2^{-n-\K(n,\mu,\lambda)-c})\geq 1/2.$$
	Let $\lambda^{n,c}$ be the probability of an encoding of $2^{n+\K(n,\mu,\lambda)+c-1}$ elements each distributed according to $\lambda$. Thus
	\begin{align*} \lambda^{n,c}(\textrm{Encoding of $2^{n+\K(n,\mu,\lambda)+c-1}$ elements $\beta$, each having $\D(\beta|\mu)\leq n$})&\geq 1/2.
	\end{align*}
	Let $v$ be a shortest program to compute $\langle n,\mu,\lambda\rangle$. By Theorem \ref{thr:coninfo}, with the universal Turing machine relativized to $v$,  $$\lambda^{n,c}(\{\gamma:\I(\gamma:\ch|v)>m\})\lem 2^{-m+\K(\lambda^{n,c}|v)}\lem 2^{-m+\K(n,\K(n,\mu,\lambda),c,\lambda|v)}\lem 2^{-m+\K(c)}.$$ Therefore, 
	\begin{align*} \lambda^{n,c}(\{\gamma:\I(\gamma:\ch|v)>\K(c)+O(1)\})&\leq 1/4.
	\end{align*}
	Thus, by probabilistic arguments, there exists $\alpha\in\IS$, such that $\alpha=\langle D\rangle$ is an encoding of $2^{n+\K(n,\mu,\lambda)+c-1}$ elements $\beta\in D\subset\IS$, where each $\beta$ has $\D(\beta|\mu)\leq n$ and $\I(\alpha:\ch|v)\lea \K(c)$. By Theorem \ref{thr:InfExt}, relativized to $v$, there are constants $d,f\in\N$ where
	\begin{align}
\nonumber
	m=\log |D|&< \max_{\beta\in D}\D(\beta|\mu,v)+2\I(D:\ch|v)+d\K(m|v)+f\K(\mu|v)\\
	m&< \max_{\beta\in D}\D(\beta|\mu)+\K(v)+2\I(D:\ch|v)+d\K(m|v)+f\K(\mu|v)\\
	\nonumber
	 &\lea \max_{\beta\in D}\D(\beta|\mu)+\K(n,\mu,\lambda)+2\K(c)+d\K(m|v)+f\K(\mu|v)\\
	 	\label{eq:h}
	&\lea n+\K(n,\mu,\lambda)+d\K(m|v)+2\K(c).		
	\end{align}
	Therefore:
	\begin{align*}
	m&= n+\K(n,\mu,\lambda)+ c-1\\
	\K(m|v)&\lea \K(c).
	\end{align*}  Plugging the inequality for $\K(m|v)$ back into Equation \ref{eq:h} results in
	\begin{align*}
	n+\K(n,\mu,\lambda)+c &\lea n+\K(n,\mu,\lambda)+2\K(c)+d\K(c)\\
	c&\lea (2+d)\K(c).
	\end{align*}
	
	This result is a contradiction for sufficiently large $c$ solely dependent on the universal Turing machine.\qed
\end{prf}\\

Similar to the construction in the introduction, we can define a universal conditional lower computable integral test $T(\alpha|n)$ over a sequence of uniformly computable measures $Q_1$, $Q_2$, $\dots$ over $\IS$. We can also define the randomness deficiency to be $\D_n(\alpha|Q)=\log T(\alpha|n)$. The following corollary is derived from the fact that $\D_n(\alpha|\mu,n)=\D_n(\alpha|\mu)$.
\begin{cor}
	For uniformly computable measures  $\{\mu_i\}$ and  nonatomic  $\{\lambda_i\}$ over $\IS$, for all $n$, $\lambda_n\{\alpha :\D_n(\alpha|\mu)>n\}> 2^{-n-\K(\mu,\lambda)-O(1)}$. 
\end{cor}

Theorem \ref{thr:dyn} can be extended to incomputable $\lambda$, which can be accomplished using a stronger version of Theorem \ref{thr:coninfo}. The term $\langle \lambda\rangle\in\IS$ in the following corollary represents any encoding of $\lambda$ that can compute $\lambda(x\IS)$ for $x\in\FS$ up to arbitrary precision. Let $\I(\lambda:\ch) = \inf_{\langle\lambda\rangle}\I(\langle\lambda\rangle;\ch)$.
\begin{cor}$ $\\ \vspace*{-0.6cm}
\begin{itemize}
\item For  measures $\mu$ and $\lambda$ over $\IS$, nonatomic $\lambda$, computable $\mu$, for all $n$, \\$\lambda\{\alpha:\D(\alpha|\mu)>n\}>2^{-n-O(\K(n,\mu)+\I( \lambda :\ch))}$.
\item For  measures $\mu$ and $\lambda$ over $\IS$, nonatomic $\lambda$, computable $\mu$, if for every $c\in\N$, there is an $n\in\N$, where $\lambda\{\alpha:\D(\alpha|\mu)>n\}<2^{-n-O(\K(n))-c}$, then $\I( \lambda:\ch)=\infty$.
\end{itemize}
\end{cor}

We define a metric $g$ on $\IS$ with $g(\alpha,\beta)=1/2^k$, where $k$ is the first place where $\alpha$ and $\beta$ disagree. Let $\mathfrak{F}$ be the topology induced by $g$ on $\IS$; $\mathcal{B}(\mathfrak{F})$ be the Borel $\sigma$-algebra on $\IS$; $\lambda$ and $\mu$ be computable measures over $\IS$ and $\lambda$ be nonatomic; and $(\IS,\mathcal{B}(\mathfrak{F}),\lambda)$ be a measure space  and $T:\IS\rightarrow\IS$ be an ergodic measure preserving transformation. By the Birkoff theorem,
\begin{cor}
\textit{Starting $\lambda$-almost everywhere, $\gem \m(n,\mu,\lambda)2^{-n}$ states $\alpha$ visited by iterations of $T$ have $\D(\alpha|\mu)>n$.}
\end{cor}
\section{Outlier Theorem}
A sampling method $A$ is a probabilistic function that maps an integer $N$ with probability 1 to a set containing $N$ different strings.
\begin{lmm}
	\label{lmm:S}
	Let $P$ be a computable measure on strings and let $A$ be a sampling method. For all integers $M$ and $N$, there exists a finite set $S\subset\FS$ such that $P(S) \leq  2M/N$, and with probability strictly more than $1-2e^{-M}$: $A(N)$ intersects $S$.
\end{lmm}
\begin{prf}
We show that some possibly infinite set S satisfies the conditions, and thus, some finite subset also satisfies the conditions due to the strict inequality. We use the probabilistic method: we select each string to be in $S$ with probability $M/N$ and show that 2 conditions are satisfied with positive probability. The expected value of $P(S)$ is $M/N$. By the Markov inequality, the probability that $P(S) > 2M/N$ is at most $1/2$. For any set $D$ containing $N$ strings, the probability that $S$ is disjoint from $D$ is
$$(1-M/N)^N<e^{-M}.$$
Let $Q$ be the measure over $N$-element sets of strings generated by the sampling algorithm $A(N)$. The left-hand side above is equal to the expected value of
	$$Q(\{D : D\textrm{ is disjoint from }S\}).$$
Again by the Markov inequality, with probability greater than $1/2$, this measure is less than $2e^{-M}$. By the union bound, the probability that at least one of the conditions is violated is less than $1/2+1/2$. Thus, with positive probability a required set is generated, and thus such a set exists.\qed
\end{prf}
\begin{thr}
	\label{thr:outlier}
Let $P=P_1,P_2\dots$ be a uniformly computable sequence of measures on strings and let $A$ be a sampling method. There exists $c\in\N$ such that for all $n$ and $k$:
	$$
	\Pr\left(\max_{a\in A(2^n)}\overline{\d}_n(a|P)> n - k -c\right)\geq 1-2e^{-2^k}.
	$$
\end{thr}
\begin{prf}
We now fix a search procedure that on input $N$ and $M$ finds a set $S_{N,M}$ that satisfies the conditions of Lemma \ref{lmm:S}. Let $t'(a|n)$ be the maximal value of $2^n/2^{k+2}$ such that $a\in S_{2^n,2^k}$ for some integer $k$. By construction, $t'$ is a computable probability bound test, because $P(\{x:t'(x|n)=2^\ell\})\leq 2^{-\ell-1}$, and thus $P(t'(x|n)\geq 2^\ell)\leq 2^{-\ell-1}+ 2^{-\ell-2}+\dots$ With the given probability, the set $A(2^n)$ intersects $S_{2^n,2^k}$. For any number $a$ in the intersection, we have $t'(x|n)\geq 2^{n-k-2}$, thus by the optimality of $t$ and definition of $\overline{\d}$, we have $\overline{\d}_n(a|P)>n-k-O(1)$.\qed
\end{prf}\\ 

An incomplete sampling method $A$ takes in a natural number $N$ and outputs, with probability $f(N)$, a set of $N$ numbers. Otherwise $A$ outputs $\perp$. $f$ is computable.
\begin{cor}
	\label{cor:inout}
Let $P=P_1,P_2\dots$ be a uniformly computable sequence of measures on strings and let $A$ be an incomplete sampling method. There exists $c\in\N$ such that for all $n$ and $k$:
	$$
	\Pr_{D=A(n)}\left(D\neq \perp\textrm{ and }\max_{a\in D}\overline{\d}_n(a|P)\leq n - k -c\right)< 2e^{-2^k}.
	$$
\end{cor}
\subsection{Continuous Sampling Method} Let $\mu=\mu_1,\mu_2,\dots$ be a uniformly computable sequence of measures over infinite sequences. Similar way as for strings in the introduction, the randomness deficiency $\overline{\D}_n(\omega|\mu)$ for sequences $\omega$ is defined using lower-semicomputable functions $\IS\times\N\rightarrow\R_{\geq 0}$. A continuous sampling method $C$ is a probabilistic function that maps, with probability 1, an integer $N$ to an infinite encoding of $N$ different sequences.

\begin{thr}
\label{thr:contsamp}
	There exists $c\in\N$ where for all $n$:
	$$
	\Pr\left(\max_{\alpha\in C(2^n)}\overline{\D}_n(\alpha|\mu)>n-k-c\right)\geq 1-2.5e^{-2^k}.
	$$
\end{thr}

\begin{prf}
For $D\subseteq\IS$, $D_m=\{\omega[0..m]:\omega\in D\}$.
Let $g(n)=\arg\min_m\Pr_{D=C(n)}(|D_m|<n)<0.5e^{-2^n}$ be the smallest number $m$ such that the initial $m$-segment of $C(n)$ are sets of $n$ strings with very high probability. $g$ is computable, because $C$ outputs a set of distinct infinite sequences with probability 1. For probability $\psi$ over $\IS$, let $\psi^m(x) = [|x|=m]\psi(\{\omega : x\sqsubset \omega\})$.
Let $\mu^g = \mu_1^{g(1)}, \mu_2^{g(2)},\dots$ be a uniformly computable sequence of discrete probability measures and let $A$ be a discrete incomplete sampling method, where for random seed $\omega\in\IS$, $A(n,\omega)=C(n,\omega)_{g(n)}$ if $|C(n,\omega)_{g(n)}|=n$; otherwise $A(n,\omega)=\perp$. So $\Pr[A(n)=\perp]<0.5e^{-2^n}$. 
\begin{align}
\nonumber
&\Pr\left(\max_{\alpha\in C(2^n)}\overline{\D}_n(\alpha|\mu)\leq n-k-O(1)\right)\\
\nonumber
\leq&\Pr_{Z= C(2^n)}\left((|Z_{g(n)}|<2^n)\textrm{ or }(|Z_{g(n)}|=2^n\textrm{ and }\max_{\alpha\in Z}\overline{\D}_n(\alpha|\mu)\leq n-k-O(1)\right)\\
\nonumber
\leq&\Pr_{D= A(2^n)}\left(D=\perp\textrm{ or }(D\neq\perp\textrm{ and }\max_{x\in D}\overline{\d}_n(x|\mu^g)\leq n-k-O(1))\right)\\
\label{eq:applysamp}
< &0.5e^{-2^n}+2e^{-2^k}\\
\nonumber
\leq& 2.5e^{-2^k},
\end{align}
where Equation \ref{eq:applysamp} is due to Corollary \ref{cor:inout}.
\qed
\end{prf}
\subsection{Alternative Proof to Theorem 3}
Using the theorem of the previous section, one can produce a simple proof to a variant of Theorem \ref{thr:dyn}. The longer proof was included due to of its tight error terms as well as its corollaries extending the results to incomputable measures. Let $\lambda = \lambda_1,\lambda_2,\dots$ and $\mu = \mu_1,\mu_2,\dots$ be uniformly computable sequences of measures over infinite sequences. Each $\lambda_n$ is non-atomic.
\begin{thr}
There are constants $b,c\in\N$, dependent on $\mu$ and $\lambda$, where for all $n\in\N$, \\$\lambda_n\left\{\alpha : \overline{\D}_n(\alpha|\mu)>n-b\right\}>2^{-n-c}$.
\end{thr}
\begin{prf} We define the continuous sampling method $C$, where on input $n$, randomly samples $n$ elements from $\lambda_n$. 
Let $d_n = \lambda_n\{\alpha :\overline{\D}_n(\alpha|\mu)>n-b\}$, where $b$ is the constant in \ref{thr:contsamp}. Evoking this theorem, with $k=0$,
\begin{align*}
 \Pr\left(\max_{\alpha\in C(2^n)}\overline{\D}_n(\alpha|\mu)>n-b\right) >& 1- 2.5e^{-1}\\
 1-(1-d_n)^{2^n} >& 1-2.5e^{-1}\\
 1-2^nd_n <& 2.5/e\\
 d_n >&(1-2.5/e)2^{-n}\\
 \lambda_n\{\alpha :\overline{\D}_n(\alpha|\mu)>n-b\} >&2^{-n-c}.
\end{align*}\qed
\end{prf}

\subsection{Necessity of Double Exponential}
Theorem \ref{thr:outlier} showed that the probability that $A(2^n)$ contains no strings of randomness deficiency less than $n-k$ decreases double exponentially in $k$. We show that at least a double exponential probability is required for $k=n-O(1)$. Let $P_n$ be the uniform measure on $(n+2)$-bit strings. The algorithm $A$ that on input $2^n$ generates a random set of $2^n$ strings of length $n+2$ satisfies
$$\Pr\left(\forall x\in A(2^n):\overline{\d}_n(x|P)\leq 2\right)\geq 2^{-2^n}.$$
The reasoning for this is as follows. For at most a quarter of the $(n+2)$-bit strings, we have $\overline{\d}_n(x|P)\geq 3$, by definition of a probability bounded test $t$. A random selection of $N=2^n$ different $(n+2)$-bit strings, contains no such string with a probability of at least $2^{-N}$. We consider the following situation. In a bag with $4N$ balls, $N$ balls are marked. One selects $N$ balls one by one. We consider the probability that no marked ball is drawn if previously no marked ball was drawn.  The smallest probability appears at the last draw when there are $T=4N-(N-1)$ balls in the bag. This probability is $(T-N)/T\geq 1/2$. 
\subsection{Partial Sampling Methods}
A partial sampling method is a sampling method that can output with  probability less than 1.
Theorem \ref{thr:outlier} does not hold for partial sampling methods $B$. Let $P_n$ be the uniform measure on $(n+1)$-bit strings. Let $\#B(N)$ represent the event that $B$ halts and outputs a set of size $N$. We present a partial  sampling method $B$ for which
$$
\Pr\left(\#B(2^n)\textrm{ and }\forall x\in B(2^n):\overline{\d}_n(x|P)\leq 1\right)\geq 2^{-n}.
$$
For at most half of the $(n+1)$-bit strings, we have $\overline{\d}_n(x|P)\geq 2$. On input $2^n$,the  partial sampling method $B$ generates a random natural number $s$ bounded by $2^n$, searches for $s$ strings $x$ of length $n+1$ with $\overline{\d}_n(x|P)\geq 2$, and outputs $2^n$ other $(n+1)$-bit strings. For some $s$, this search may never terminate. If $A$ chooses to be precisely equal to the number of strings satisfying the condition, then it outputs only strings with deficiency at most $1$, and the claim is proven. However partial sampling methods do exhibit the following properties
\begin{thr}
	Let $P=P_1,P_2,\dots$ be a uniformly computable sequence of measures and $B$ be a partial sampling method, where $\#B(N)$ represents the event that $B(N)$ terminates and outputs a set of $N$ strings.
	$$
	\Pr\left(\#B(N)\textrm{ and }\forall x\in B(2^n):\overline{\d}_n(x|P)\leq n-k\right)\leq O(k2^{-k}).
	$$
\end{thr}
\begin{prf}
	Let $Q$ be the lower-semicomputable semimeasure over sets of size $2^n$ such that $Q(D)$ equals the probability that $B(N)=D$. We show that
		$$
	\Pr\left(\#B(N)\textrm{ and }\forall x\in B(2^n):\overline{\d}_n(x|P)\leq n-k+\log k+O(1)\right)\leq O(2^{-k}).
	$$
	This result is  followed by a redefinition of $k$. We write $Q$ as a uniform mixture over at most $2^k$ measures $Q_i$ with finite support, and one lower semi-computable semimeasure $Q_*$:
	$$
	Q=2^{-k}\left(Q_1+Q_2+\dots Q_f+Q_*\right).
	$$
	With $f\leq 2^k$, we assume that the finite descriptions of $Q_1,\dots,Q_f$ are enumerated one by one by a program (that may never terminate). For each enumerated measure $Q$, we search for a set $S_i$ that satisfies the conditions of Lemma \ref{lmm:S} for $M=k$. Let $S=\bigcup_{i\leq f}S_i$. Also, $P(S)\leq k2^{k+1-n}$; thus every element in $S$ satisfies $\overline{\d}_n(x|P)\geq n-k+\log k+O(1)$. 
	
	The probability that $A(2^n)$ produces a set that does not contain such an element is at most $2^{-k}+2e^{-k}$ because we can equivalently generate a set $D$ by randomly selecting $j$ from the list $[1,\dots,f,*,\infty]$ with probabilities $[2^{-k},\dots,2^{-k},2^{-k}r,1-(f+r)2^{-k}]$ and generating a random set $D$ from $Q_j$ if $j\neq\infty$ and letting $D$ be undefined otherwise. The probability that $D$ is defined and does not contain an element from $S$ is at most the probability $j=*$, which is $\leq 2^{-k}$, plus the probability that $j\in\{1,\dots,f\}$ times $2e^{-k}$.\qed
\end{prf}

\section{Function Derandomization}

In this section we show how to construct deterministic functions from random ones. The main results of this section are not (directly) implied by the main theorem in \cite{Levin16,Epstein19}, because Theorem \ref{lmm:derand} is a statement about probabilities over the Baire space, whereas the result in \cite{Levin16,Epstein19} is a statement about lower computable semi measures over $\N$. Similarly, the main results in Sections \ref{sec:open} and \ref{sec:monotone} are statements about computable and lower computable continuous semi measures over the Cantor space.

We recall the definitions from the introduction. Random functions $F$ over natural numbers are modeled by discrete stochastic processes indexed by $\N$, where each $F(t)$, $t\in\N$, is a random variable over $\N$. $\mathcal{F}$ is the set of all random functions. A 
random function $F\in\mathcal{F}$ is computable if there is a program that on input $(a_1,\dots,a_n)$ lower computes $\Pr\left[F(1)=a_1\cap F(2)=a_2\cap\dots \cap F(n)=(a_n) \right]$. Put another way, a random function $F\in\mathcal{F}$ is computable if $X=\Pr[F(a_1)=b_1 \cap\dots\cap F(a_n)=b_n]$ is uniformlly computble in $\{(a_i,b_i)\}_{i=1}^n$.  
 The complexity $\K(F)$ of a random function $F\in\mathcal{F}$, is the smallest program that
computes $X$. $\mathcal{G}$ is the set of all deterministic functions $G:\N\rightarrow\N$. A sample $S\in\mathcal{S}$ is a finite set of pairs $\{(a_i,b_i)\}_{i=1}^n$. The encoding of a sample is $\langle S\rangle= \langle \{(a_i,b_i\}_{i=1}^n\rangle$. $\mathcal{S}$ is the set of all samples. We say $G(S)$ if $G$ is consistent with $S$, with $G(a_i)=b_i$, $i=1,\dots,n$. For random functions, $F(S)$ is the event that $F$ is consistent with $S$.

To prove function derandomization, we leverage properties about the Baire space $\N^\N$.
 Individual cylinders are $C_n[v] = \{(a_1,a_2,\dots)\in\N^\N:a_n=v\}$. Cylinders are generators for cylinder sets. The cylinder sets $C\in\mathcal{C}$ consists of all intersections of a finite number of cylinders. If $C = \bigcap_{i\in I}C_i[v_i]$, then for all $i\in I$, we say $i\in\mathrm{Dom}(C)$. The set of all such cylinder sets provides a basis for the product topology of $\N^\N$. The encoding of a cylinder set $C = \bigcap_{i\in I}C_i[v_i]$, is $\langle C\rangle = \langle \{i,v_i\}_{i\in I}\rangle$.  The set of all Borel probability measures over $\N^\N$ is $\mathcal{P}$. A probability $P\in\mathcal{P}$ is computable if given an encoding
of a cylinder set $C\in\mathcal{C}$, $P(C)$ is computable. 
 
 We use the following helper proposition and lemma throughout the paper.
 \begin{prp}$ $\\
	\label{prp:monotonic}
	For every $c,n\in\N$, if $x<y+c$ for some $x,y\in\N$m then $x+n\K(x)< y+n\K(y)+O(n\log n)+2c$.
\end{prp}
\begin{prf}
	$\K(x)\lea \K(y)+\K(y-x)$ as $x$ can be computed from $y$ and $(y-x)$. Therefore $n\K(x)-n\K(y) < n\K(y-x)+dn$, for some $d\in\N$ dependent on $U$. We assume that this equation is not true; then, there exists $x,y,c\in\N$ where $x<y+c$, and $g\leq O(n\log n)+2c$ where
	 $y-x+g< n\K(x)-n\K(y)<n\K(y-x)+dn$, which is a contradiction for $g\eqa dn+2c+\max_a
	\{2n\log a-a\}\eqa dn+2c+2n\log n$.\qed
\end{prf}

\begin{lmm}
	\label{lmm:consH}
		For partial computable $f:\N\rightarrow\N$, for all $a\in\N$, $\I(f(a);\mathcal{H})\lea\I(a;\mathcal{H})+\K(f)$. 
\end{lmm}
\begin{prf}
	\begin{align*}
	\I(a;\ch)&=\K(a)-\K(a|\ch)\gea \K(a,f(a))-\K(a,f(a)|\ch)-\K(f).
	\end{align*}
	The chain rule ($\K(x,y)\eqa \K(x)+\K(y|x,\K(x))$) applied twice results in
	\begin{align*}
	\I(a;\ch)+\K(f)&\gea \K(f(a))+\K(a|f(a),\K(f(a)))-(\K(f(a)|\ch)+\K(a|f(a),\K(f(a)|\ch),\ch)\\\
	&\eqa \I(f(a);\ch)+\K(a|f(a),\K(f(a)))-\K(a|f(a),\K(f(a)|\ch),\ch)\\
	&\eqa  \I(f(a);\ch)+\K(a|f(a),\K(f(a)))-\K(a|f(a),\K(f(a)),\K(f(a)|\ch),\ch)\\
	&\gea \I(f(a);\ch).
	\end{align*}
	\qed
\end{prf}

 \begin{thr}
 \label{thr:samp}
 For $F\in\mathcal{F}$, $S\in\mathcal{S}$, if $s=\ceil{-\log \Pr[F(S)]}$ and $h=\I(\langle S\rangle;\mathcal{H})$, then \\
 $\min_{G\in\mathcal{G},G(S)}\K(G) < \K(F)+s+h+O(\K(s,h)+\log\K(F))$.
 \end{thr}
 \begin{prf}

Each sample $S\in\mathcal{S}$ where $S=\{(i,v_i)\}_{i\in I}$ can be identified by a cylinder set $C_S\in\mathcal{C}$ where $C_S = \cap_{i\in I}C_i[v_i]$. For every $\alpha \in \N^\N$ there is a deterministic function $G_\alpha:\N\rightarrow\N$, where $G_\alpha(i)=\alpha[i]$. Furthermore if $\alpha \in C_S$, then for all $(i,v_i)\in S$, $G_\alpha(i)=v_i$. For each random function $F\in\mathcal{F}$, we can identify a Borel probability $P_F\in\mathcal{P}$ over $\N^\N$ such that for each sample $S=\{(i,v_i)\}_{i\in I}\in\mathcal{S}$,  $\Pr[F(S)]=P_F(C_S)$. This is because random functions and Borel probability measures over $\N^\N$ have the same form. Furthermore, if $F$ is computable, then $P_F$ is computable, with 
\begin{align}
\label{eq:pff}
\K(P_F|F) &= O(1)
\end{align}
This is because given an encoding $\langle F\rangle$ and an encoded cylinder set $\langle C\rangle$, one can compute $\Pr[F(C)]$, which is equal to $P_F(C)$. Thus given a random function $F\in\mathcal{F}$ and sample $S\in\mathcal{S}$, by Lemma \ref{lmm:derand} applied to $P_F\in\mathcal{P}$ and $C_S\in\mathcal{C}$, we get the following result, with $h_C=\I(\langle C_S\rangle;\ch)$, $h_S = \I(\langle S\rangle;\ch)$, and $s = \ceil{-\log P_F(C_S)}$,
\begin{align}
\nonumber
\min_{\alpha \in C_S}\K(a) <& \K(P_F) +s+ h_C + O(\K(s)+\log \K(P_F)) + O(\K(h_C))\\
\label{eq:alphaG}
\min_{G\in \mathcal{G}:G(S)}\K(G) <& \K(P_F) +s + h_C + O(\K(s)+\log\K(P_F)) + O(\K(h_C))\\
\label{eq:applyKPFF}
\min_{G\in \mathcal{G}:G(S)}\K(G) <& \K(F) +s+ h_C + O(\K(s)+\log\K(F)) + O(\K(h_C))\\
\label{eq:applycons}
\min_{G\in\mathcal{G}:G(S)}\K(G) <& \K(F) +s+ h_S + O(\K(s)+\log\K(F)) + O(\K(h_S))\\
\label{eq:defcs}
\min_{G\in\mathcal{G},G(S)}\K(G) <& \K(F) -\log\Pr\left[F(S)\right] + \I(\langle S\rangle;\ch)\\
\nonumber
& + O(\K(\ceil{-\log\Pr\left[F(S)\right]},\I(\langle S\rangle;\ch))+\log \K(F)).
\end{align}
Equation \ref{eq:alphaG} is because for the $\alpha\in\N^\N$ that minimizes the leftmost term, $G_\alpha \in\mathcal{G}$, with $G_\alpha(S)$ and $\K(G_\alpha)\lea \K(\alpha)$.
Equation \ref{eq:applyKPFF} is because $P_F$ can be constructed from $F$, i.e. Equation \ref{eq:pff}.
Equation \ref{eq:applycons} is due to Proposition \ref{prp:monotonic}, Lemma \ref{lmm:consH} and the fact that $\K(\langle C_S\rangle|\langle S\rangle)=O(1)$. Equation \ref{eq:defcs} is due to the definition of $s$,  where $s = \ceil{-\log P_F(C_S)}=\ceil{-\log \Pr[F(S)]}$.
\qed
\end{prf}

\begin{lmm}
\label{lmm:derand}
For cylinder set $C\in\mathcal{C}$, computable probability $P\in\mathcal{P}$, if $s=\ceil{-\log P(C)}$ and $h=\I(\langle C\rangle;\ch)$,  then $\min_{\alpha\in C}\K(\alpha)< \K(P)+s+h+O(\K(s,h)+\log \K(P))$.
\end{lmm}

\begin{prf}  We put $(s,P)$ on an auxiliary tape to the universal Turing machine $U$. Thus, all algorithms have access to $(s,P)$, and all complexities implicitly have $(s,P)$ as conditional terms. 

Let $Q$ be an elementary probability measure that realizes $\Ks(\langle C\rangle)$. Let $d=\max\{\d(\langle C\rangle|Q),1\}$ and $c\in\N$ be a constant to be chosen later. Let $n=\max \{m:m\in\mathrm{Dom}(W), W\in\mathcal{C}, \langle W\rangle\in\mathrm{Supp}(Q)\}$. For a list $L$ of a list of numbers and cylinder set $W\in\mathcal{C}$, we say $L\rtimes W$ is the set of all $x\in L$ with $x\N^\N\subseteq W$. We define a measure $\kappa$ over $cd2^s$ lists of lists of $n$ numbers $L$, where $\kappa(L) = \prod_{i=1}^{cd2^s}P(L[i]\N^\N)$. Given a list of lists of $n$ numbers $L$, $\kappa(L)$ is computable (as a program for $P$ is on an auxiliary tape). We use the indicator function  $\i(L,W) = [W\in\mathcal{C}, P(W)\geq 2^{-s}, L \rtimes W=\emptyset]$. The function $\i$ is computable, because $P(W)$ and $L\rtimes W$ are computable for all $W\in\mathcal{C}$.
\begin{align*}
\E_{L\sim\kappa}\E_{W\sim Q}[\i(L,W)]&\leq\sum_WQ(W)\Pr_{L\sim\kappa}(W\in\mathcal{C}, P(W)\geq 2^{-s}, L \rtimes W=\emptyset)\\
&\leq\sum_WQ(W)\prod_{i=1}^{cd2^s}(1-2^{-s})\\
&\leq \sum_WQ(W)(1-2^{-s})^{cd2^s}\\
&<e^{-cd}.
\end{align*}
Thus there exists a list $L'$ of $cd2^s$ sequences of numbers of length $n$ such that $\E_{W\sim Q}[\i(W,L')]=e^{-cd}$. Thus $t(W) = \i(W,L')e^{cd}$ is a $Q$-test, with $t:\FS\rightarrow\R_{\geq 0}$  and $\sum_WQ(W)t(W)\leq 1$. It must be that $L\rtimes C\neq\emptyset$. Otherwise $t(C)=e^{cd}$, and 

\begin{align*}
\K(C|c,d,Q) &\lea -\log t(C)Q(C)\\
&\lea -\log Q(C) - (\lg e)cd\\
(\lg e)cd &\lea-\log Q(C) - \K(C|P) + \K(d,c)\\
(\lg e)cd&< d + \K(d,c) + O(1).
\end{align*}

This is a contradiction for $c$ large enough solely dependent on the universal Turing machine. We roll $c$ into the additive constants of the rest of the proof. Thus there exists $x\in L \rtimes C$ where 
\begin{align}
\nonumber
\K(x) &\lea \log |L| +\K(L)\\
\nonumber
&\lea \log |L| +\K(d,Q)\\
\nonumber
&\lea \log d + s + \K(d)+\K(Q)\\
\nonumber
&\lea s +3\log d +\K(Q)\\
\label{eq:applystoch}
&\lea s+\Ks(C)\\
\label{eq:alphax}
\min_{\alpha\in C}\K(\alpha)\lea \K(x) &\lea s +\Ks(C),
\end{align}
where Equation \ref{eq:applystoch} is due to the definition of stochasticity. Equation \ref{eq:alphax} is because $x\N^\N\subseteq C$. Thus making the relativization of $(s,P)$ explicit,
\begin{align}
\nonumber
\min_{\alpha\in C}\K(\alpha|\langle s,P\rangle)&\lea s +\Ks(\langle C\rangle|\langle s,P\rangle)\\
\nonumber
\min_{\alpha\in C}\K(\alpha)&< \K(P)+s +\Ks(\langle C\rangle)+O(\K(s)+\log \K(P))\\
\label{eq:applystochh}
\min_{\alpha\in C}\K(\alpha)&< \K(P)+s +\I(\langle C\rangle;\ch)+O(\K(s,\I(\langle C\rangle;\ch))+\log \K(P)).
\end{align}
Equation \ref{eq:applystochh} follows from Lemma 10 in \cite{Epstein21}, which states $\Ks(x)<\I(x;\mathcal{H})+O(\K(\I(x;\mathcal{H})))$.\qed
\end{prf}$ $\\

Theorem \ref{thr:samp} can be readily extended to sets of samples $\mathfrak{S} = \{S_1,\dots,S_n\}$, where for deterministic function $G:\N\rightarrow\N$, $G(\mathfrak{S})$ if $\bigcup_{i=1}^nG(S_i)$. For random function $F\in\mathcal{F}$, $F(\mathfrak{S})$ is the union of events $F(S_i)$, $i=1,\dots,n$. The proof of the following corollary follows almost identically to the proofs of Theorem \ref{thr:samp} and Lemma \ref{lmm:derand}, noting that $P(\mathfrak{S})$ is computable given a computable probability $P\in\mathcal{P}$ and a finite description of a set of samples $\mathfrak{S}$.
\begin{cor}
\label{cor:setsample}
 For $F\in\mathcal{F}$, if $s=\ceil{-\log \Pr[F(\mathfrak{S}
 )]}$ and $h=\I(\langle \mathfrak{S}\rangle;\mathcal{H})$, then 
 $\min_{G\in\mathcal{G},G(\mathfrak{S})}\K(G) \lel \K(F)  +s+h+O(\K(s,h)+\log\K(F))$.
\end{cor}

Another generalization of Theorem \ref{thr:samp} is in the usage of lower computable random functions $V$. They are discrete stochastic processes $V(t)$, indexed by $t \in\N$, where each $V(t)$ is a random  variable over $\N \cup \infty$. Furthermore $\Pr(V(1)=a_1 \cap V(2) = a_2\cap\dots \cap V(n)=a_n)$ is lower computable, where $a_i \in \N$, $i=1\dots n$. The proof is extensive, relying on left total machines, introduced in \cite{Levin16,Epstein19}. It is not included in this paper.

\begin{exm}[Classification]
Lets say we have $m$ disjoint groups $J_i\subset\N$, where $|J_i|=n$, for each $i$. We want to find the simpliest total function $G:\N\rightarrow \N$ such that $J_i\subseteq G^{-1}(i)$ for each $i$. We can construct a random function $F\in\mathcal{F}$, where $\Pr[F(a)=i]=1/m$, for $i=1,\dots,m$, for all $a\in\N$. Thus for the event $F(E)$ that $F$ produces $i$ over each $J_i$, $\Pr[F(E)]=\prod_{i=1}^m(1/m)^n=m^{-mn}$. So by Corollary \ref{cor:setsample}, there is a deterministic function $G$ such that $J_i\subseteq G^{-1}(i)$ and $\K(G) \lel \K(F) -\log \Pr[F(E)] + \I(\langle \{J_i\}\rangle;\ch)\lel \K(F) -\log m^{-mn} + \I(\langle \{J_i\}\rangle;\ch)\lel \K(m) +mn\log m + \I(\langle \{J_i\}\rangle;\ch)$.
\end{exm}

\section{Open Sets}
\label{sec:open} 
We recall that the Kolmogorov complexity of an infinite sequence $\alpha\in\IS$ is $\K(\alpha)$, the size of the smallest program to a universal Turing machine that will output, without halting, $\alpha$ on the output tape.

\begin{thr}
\label{thr:clopen}
For clopen set $C\subseteq\IS$, if $s=\ceil{-\log\mu(C)}$ and $h=\I(\langle C\rangle;\ch)$, then \\$\min_{\alpha\in C}\K(\alpha)< s+h + O(\K(s,h))$.
\end{thr}
\begin{prf}
We define a set of samples $\mathfrak{S}$, where for each maximal interval $\Gamma_x\subseteq C$, $x\in\FS$, we add the sample $S=\{(i,x[i])\}_{i=1}^{\|x\|}$ to $\mathfrak{S}$. Thus $\K(\langle \mathfrak{S}\rangle|\langle C\rangle)=O(1)$. Furthermore we define a stochastic process $F(t)$ over $\N$, indexed by $t\in\N$ using the uniform distribution $\mu$ over $\IS$, where $\Pr[F(1)=a_1,F(2)=a_2,\dots,F(n)=a_n]=2^{-n}\prod_{i=1}^n[a_i\in\BT]$. Thus $s=\ceil{-\log\Pr[F(\mathfrak{S})]}=\ceil{-\log \mu(C)}$. Using Corollary \ref{cor:setsample}, noting that $\K(F)=O(1)$,
\begin{align}
\nonumber
\min_{G\in\mathcal{G},G(\mathfrak{S})}\K(G)&< s + \I(\langle \mathfrak{S}\rangle;\ch)+O(\K(s))+O(\K(\I(\langle \mathfrak{S}\rangle;\ch)))\\ 
\label{eq:clopenapplycons}
\min_{G\in\mathcal{G},G(\mathfrak{S})}\K(G)&< s + \I(\langle C\rangle;\ch)+O(\K(s))+O(\K(\I(\langle C\rangle;\ch)))\\
\label{eq:Gtoa}
\min_{\alpha \in C}\K(\alpha)&< s + \I(\langle C\rangle;\ch)+O(\K(s,\I(\langle C\rangle;\ch))).
\end{align}
 Equation \ref{eq:clopenapplycons} is due to Lemma \ref{lmm:consH} and Proposition \ref{prp:monotonic}, noting that $\K(\langle \mathfrak{S}\rangle|\langle C\rangle)=O(1)$. Equation \ref{eq:Gtoa} comes from modifying $G$ to having it output 0 whenever it would normally output a number $b\not\in \BT$. This new function $\alpha$ can be thought of an infinite sequence $\alpha\in\IS$, and since $G(\mathfrak{S})$, it must be that $\alpha\in C$. Furthermore $\K(\alpha|G)=O(1)$.\qed
 \end{prf}
 
 \begin{exm}[{\sc Max-3Sat}]
 \label{exm:sat}
 This problem consists of a boolean formula $f$ in conjunctive normal form, comprised of $m$ clauses, each consisting of a disjunction of 3 literals. Each literal is either a variable or the negation of a variable. We assume that no literal (including its negation) appears more than once in the same clause. There are $n$ variables. The goal is to find an assignment of variables that satisfies as many clauses as possixble. The randomized approximation algorithm is as follows. The variables are assigned true or false with equal probability. Let $Y_i$ be the random variable that clause $i$ is satisfied. Thus the probability that clause $Y_i$ is satisfied is $7/8$. So the total expected number of satisfied clauses is $7m/8$, which is $7/8$ of optimal. Some simple math shows the probability that number of satified clauses is $>6m/7$ is at least $1/8$.
 
We can model this randomized algorithm using a clopen set. Let $x\in\BT^n$ encode an assignment of $n$ variables, where $x[i]=1$ if variable $i$ is true. Let clopen set $C$ be equal to $\bigcup\{\Gamma_x:\textrm{ $x$ encodes an assignment where $> 6m/7$ clauses are satisfied}\}$. The randomized assignment algorithm is modeled by the uniform measure, with $\mu(C)\geq 1/8$. Furthermore $\K(\langle C\rangle |f)=O(1)$. By Theorem \ref{thr:clopen} and Lemma \ref{lmm:consH},
$$
\min_{\alpha\in C}\K(\alpha)\lel -\log \mu(C) + \I(\langle C\rangle;\ch)\lel \I(f;\ch).
$$
Thus there is a total algorithm that can assign variables to satisfy $6/7$ the optimal number of satisfied clauses. This algorithm has complexity $\lel\I(f;\ch)$.
 \end{exm}
\begin{exm}[{\sc Max-Cut}]
\label{exm:cut}
Imagine a graph $G=(E,V)$ consisting of vertices $V$ and undirected edges $E$, and a weight $\omega_e$ for each edge $e\in \E$. Let $\omega=\sum_{e\in E}\omega_e$ be the combined weight of all edges. The goal is to find a partition $(A,B)$ of the vertices into two groups that maximizes the total weight of the edges between them. Imagine the algorithm that on receipt of a vertex, randomly places it into $A$ or $B$ with equal probability. Then the expected weight of the cut is
$$
\E\left[\sum_{e\in E(A,B)}\omega_e\right]=\sum_{e\in E}\omega_e\Pr(e\in E(A,B)) = \frac{1}{2}\omega.
$$
This means the expected weight of the cut is at least half the weight of the maximum cut. Some simple math results in the fact that $\Pr\left[\sum_{e\in E(A,B)}\omega_e\right]>\omega/3 \geq 1/4$. We can encode a cut into a binary string of $x$ length $|V|$, where $x[i]=1$, if the $i$th vertex is in $A$. The sorting algorithm is modeled with the uniform measure $\mu$. Let $C=\bigcup\{\Gamma_x:\textrm{ $x$ is an encoded cut that has combined weight $>\omega/3$}\}$. Thus $\K(C|G)=O(1)$ and $\mu(C)\geq 1/4$.
By Theorem \ref{thr:clopen} and Lemma \ref{lmm:consH},
\begin{align*}
\min_{\alpha\in C}\K(\alpha) &\lel -\log \mu(C)+\I(\langle C\rangle;\ch)\lel \I(G;\ch).
\end{align*}
Thus there is a total algorithm that can partition the vertices of graph $G$ into two groups such that the weight of its cut is 1/3 optimal. This algorithm has complexity $\lel\I(G;\mathcal{H})$.
\end{exm}
Note that another way to achieve the {\sc Max-Cut} approximation bounds is by using Theorem \ref{thr:el}, which would add an $\K(n)$ term to demarcate the number of vertices in the graph, and similarly for the {\sc Max-3Sat}  problem.
 \begin{exm}
 Let clopen set $C\subset\IS$ be defined by $\bigcup\{\Gamma_x:\|x\|=n, \K(x) > n-c\}$, for some small $c\in \N$. Thus $\ceil{-\log\mu(C)}=O(1)$ and $\min_{\alpha\in C}\K(\alpha)\gea n$ because if $\alpha\in C$, then $\alpha[0..n]$ is a random string. Furthermore $\I(\langle C\rangle;\ch)\gea n-\K(n)$ because for the first interval $\Gamma_x$ encoded in $\langle C\rangle$,  $\K(\langle C\rangle)\gea \K(\langle \Gamma_x\rangle)\gea n$, and $\K(\langle C\rangle|\ch)\lea \K(n)$.
 \end{exm}

Theorem \ref{thr:clopen} can be generalized to arbitrary open sets of the Cantor space. Such sets $S$ can have encodings $\langle S\rangle$ that are infinite sequences.  We recall that the information term between infinite sequences is $\I(\alpha:\beta)=\log\sum_{x,y\in\FS}\m(x|\alpha)\m(y|\beta)2^{\I(x:y)}$.

\begin{thr}
\label{thr:open}
		For open set $S\subseteq\IS$, if $s=\ceil{-\log\mu(S)}$ and $h=\I(\langle S\rangle:\mathcal{H})$, then
	$ \min_{\alpha\in S}\K(\alpha)<s+h+O(\K(s,h))$.
\end{thr}
\begin{prf}
	Let $\{x_i\}_{i=1}^n=\{x:\Gamma_x\textrm{ is maximal in } S\}$, with $n\in\N\cup\infty$. Let $N\in \N$ be the smallest number such that $\sum_{i=1}^N2^{-\|x_i\|}>2^{-s-1}$. Let $C=\bigcup_{i=1}^N\Gamma_{x_i}$ be a clopen set with $C\subseteq S$. By Theorem \ref{thr:clopen},
	\begin{align}
	\label{eq:open1}
	\min_{\alpha \in C}\K(\alpha)&< s + \I(\langle C\rangle;\mathcal{H})+O(\K(s))+O(\K(\I(\langle C\rangle;\mathcal{H}))).
	\end{align}
	Based on the definition of $\I$:
	\begin{align*}
	\I(\langle C\rangle;\mathcal{H})&\lea \I(\langle S\rangle:\mathcal{H})+\K(\langle C\rangle|\langle S\rangle)\\
&\lea \I(\langle S\rangle:\mathcal{H})+\K(s).
	\end{align*}
	By Proposition \ref{prp:monotonic}, where $x=\I(\langle C\rangle;\mathcal{H})$, $y=\I(\langle S\rangle:\mathcal{H})$, and $c=\K(s)+O(1)$, 
\begin{align}
\label{eq:open2}
\I(\langle C\rangle;\mathcal{H})+O(\K(
\I(\langle C\rangle;\mathcal{H})))&<
\I(\langle S\rangle:\mathcal{H})+O(\K(
\I(\langle S\rangle:\mathcal{H})))+O(\K(s)).
\end{align}
Putting Equations \ref{eq:open1} and \ref{eq:open2} together results in
\begin{align*}
	\min_{\alpha \in S}\K(\alpha)&< s + \I(\langle S\rangle:\mathcal{H})+O(\K(s,\I(\langle S\rangle:\mathcal{H}))).
\end{align*}\qed
\end{prf}

\section{Closed Sets}
There is no equivalent to Theorem \ref{thr:open} for closed sets. For closed sets $S\subseteq\IS$ of infinite strings $S_{\leq n}=\{\alpha[0..n]:\alpha \in S\}$ and $\langle S\rangle =\langle S_{\leq 1}\rangle\langle S_{\leq 2}\rangle\langle S_{\leq 3}\rangle\dots$ The closed set theorem uses the following proposition of conservation of information with respect to a partial computable function. The complexity of a partial computable function $f$, is $\K(f)$, the minimal length of a $U$-program to compute $f$. A short proof to the following proposition can be found in \cite{Geiger12}.
\begin{prp}
\label{prp:consbi}
	For $\alpha,\beta\in\IS$, partial computable $f:\IS\rightarrow\IS$,
	$\I(f(\alpha):\beta)<\I(\alpha:\beta)+\K(f)$.
\end{prp}
\begin{thr}[Anonymous]
	There exists a closed set $C\subset\IS$ consisting of solely uncomputable sequences, $\mu(C)>0$, and $\I(\langle C\rangle:\mathcal{H})<\infty$.
\end{thr}
\begin{prf}
	Let $d$ be any positive constant and $\alpha\in\IS$ be any uncomputable sequence such that $\I(\alpha:\mathcal{H})<\infty$. We inductively define a total computable function $f$ such that $f(\alpha)=\langle C\rangle$ for some closed set $C$. At round 0, $f(a)$ outputs $\langle C_{\leq 0}\rangle=\langle\rangle$. Assume $f(\alpha)$ has outputted $\langle C_{\leq 1}\rangle\langle C_{\leq 2}\rangle\dots\langle C_{\leq n-1}\rangle$.

	Let $\K_t(x)=\inf\{\|p\|:U(p)=x\textrm{ in }\leq t\textrm{ steps}\}$. Let $S$ consist of the set $y\sqsubseteq x\in C_{\leq n-1}$ such that  $\|y\|-\K_n(y)>d$. $C_{\leq n}$ is constructed in the following way. For each $x\in C_{\leq n-1}$, if there is a $y\in S$, with $y\sqsubseteq x$, then $x(\alpha[\|x\|+1])$ is added to $C_{\leq n}$. Otherwise $x0$ and $x1$ are added to $C_{\leq n}$. The function $f$ then appends $\langle C_{\leq n}\rangle$ to the output and proceeds to step $n+1$. By Proposition \ref{prp:consbi}, the amount of mutual information that $C$ has with $\mathcal{H}$ is $\I(\langle C\rangle:\mathcal{H})\lea \I(\alpha:\mathcal{H})+\K(f)<\infty$. Furthermore $\mu(C)\geq \mu(\{\alpha : \D(\alpha)\leq d\})>0$, where $\D(\alpha)=\sup_{x\sqsubset \alpha}\|x\|-\K(x)$. Every  $\alpha \in C$ either has $\D(\alpha)\leq d$ or is equal to $x\alpha_{>\|x\|}$ for some $x\in\FS$, and is thus uncomputable.\qed
\end{prf}

\section{Algorithmic Monotone Probability of Sets}
\label{sec:monotone}
In \cite{Levin16,Epstein19}, the combined algorithmic probability $\sum_{x\in D}\m(x)$ of a non-exotic set $D$ was shown to be close to $\max_{x\in D}\m(x)$. In this section, we prove an analogous theorem with the universal lower-computable continuous semi-measure $\M$. The two results are related, but neither one is readily entailed by the other.

A continuous semi-measure $Q$ is a function $Q:\FS\rightarrow \R_{\geq 0}$, such that $Q(\emptyset) =1$ and for all $x\in\FS$, $Q(x)\geq Q(x0)+Q(x1)$. For prefix free set $D$, $Q(D)=\sum_{x\in D}Q(x)$.  Let $\M$ be a largest, up to a multiplicative factor, lower semi-computable continuous semi-measure. That is, for all lower computable continuous semi-measures $Q$ there is a constant $c\in\N$ where for all $x\in\FS$, $c\M(x)>Q(x)$. Thus for any lower computable continuous semi-measure $W$ and open set $S$, $-\log\M(S)\lea \K(W) -\log W(S)$, where $\K(W)$ is the size of the smallest program that lower computes $W$.

The monotone complexity of a finite prefix-free set $G$ of finite strings is $\Km(G)\edf \min\{\|p\|\,{:}\; U(p)\in x\sqsupseteq y\in G\}$. Note that this differs from the usual definition of $\Km$, in that our definition requires $U$ to halt. A total computable function $\nu\,{:}\,\FS\,{\rightarrow}\,\FS$ is prefix-monotonic iff for all strings $x$ and $y$,  $\nu(x)\,{\sqsubseteq}\,\nu(xy)$. Let $\overline{\nu}\,{:}\,\FS\cup\IS\,{\rightarrow}\,\FS\cup\IS$ be used to represent the unique extension of $\nu$ to infinite sequences. Its definition for all $\alpha\in\FS\cup\IS$ is $\overline{\nu}(\alpha)= \sup\, \{\nu(\alpha_{\leq n})\,{:}\,n\,{\leq}\,\|\alpha\|\}$, where the supremum is respect to the partial order derived with the $\sqsubseteq$ relation. The following theorem relates prefix monotone machines and continuous semi-measures. It is equivalent to Theorem 4.5.2 in \cite{LiVi08}, with the simple modification that the machine be total computable. 
\begin{thr}
	\label{thr:stringMonotonic}
	For each lower-computable continuous semi-measure $\sigma$ over $\IS$, there is a prefix-monotonic function $\nu_\sigma$, where for prefix free $G\subset \FS$, $\ceil{-\log \sigma(G)}{\eqa} \ceil{-\log \mu\{\alpha{:}\overline{\nu_\sigma}(\alpha)\sqsupseteq x\in G\}}$.
\end{thr}

Since there is a universal lower-semicomputable continuous semi-measure $\M$, there exists a prefix-monotonic function $\nu_\M$, with the following property.

\begin{cor}
	\label{cor:KMvM}
	For finite prefix free set $G$,
	$-\log\M(G)\,{\eqa}\, -\log \mu\{\alpha\,{:}\,x\sqsubseteq \overline{\nu_\M}(\alpha),\,\alpha\,{\in}\,\IS, x\in G\}$. 
\end{cor}
The following corollary is equivalent to Theorem \ref{thr:clopen} in terms of finite strings instead of clopen sets. For finite prefix free set $G\subset\FS$, $\mu(G) = \sum_{x\in G}2^{-\|x\|}$.

\begin{cor}
	\label{cor:clopenu}
	For finite prefix free $G\subset\FS$, $s=\ceil{-\log\mu(G)}$, and $h=\I(G;\ch)$, we have $\min_{y\sqsupseteq x\in G}\K(y)< s+h + O(\K(s,h))$.
\end{cor}

\begin{thr}
\label{thr:mon}
	For finite prefix-free set $G\subset\FS$, $i=\ceil{-\log\M(G)}$, $h=\I(G;\ch)$, we have\\ $\Km(G)< i+h+O(\K(i,h))$.
\end{thr}
\begin{prf}
	Let $F\subset\FS$ be a finite prefix-free set, such that
	\begin{enumerate}
	\item $-\log\mu(F)\leq i+1$,
	\item for all $x\in F$, $\nu_\M(x)\sqsupseteq z\in G$,
	\item  $\K(F|G)\lea \K(i)$.
	\end{enumerate}
By Corollary \ref{cor:clopenu}, there exists $y \sqsupseteq x\in  F$, with $\K(y)< i+h'+O(\K(i,h'))$, where $h'=\I(F;\mathcal{H})$. Using Proposition \ref{prp:monotonic} and Lemma \ref{lmm:consH}, $\K(y)< i+h+O(\K(i,h))$, noting that $\K(F|G)\lea \K(i)$. Thus there is a program $p$ of length $\lea \K(y)$ that computes $y$ and then outputs $\nu_\M(y)\sqsupseteq \nu_\M(x)\sqsupseteq z\in G$. So $\Km(G)\lea \|p\| \lea \K(y) < i+h+O(\K(i,h))$. \qed
\end{prf}
\begin{cor}
For (potentially infinite) prefix-free set $G\subset\FS$, where if $i=\ceil{-\log\Km(G)}$, $h=\I(\langle G\rangle:\ch)$, then $\Km(G)< i+h+O(\K(i,h))$.
\end{cor}
The proof of this corollary follows analogously to the proof of Theorem \ref{thr:open}, except $\M$ is used instead of $\mu$.

\begin{exm}
Let $S=\{(a_i,b_i)\}_{i=1}^n$ be a sample, where each $a_i\in\N$, $b_i\in\BT$. Let $p=|\{i:b_i=1\}|/n$ be the percentage of locations where the sample is $1$. Let $\mathcal{B}_p$ be the Bernoulli distribution over $\IS$ with parameter $p$. Thus, using a common bound to the binary entropy function, $-\log\mathcal{B}_p(S) = -\log\left((p)^{np}(1-p)^{n(1-p)}\right)\leq n(-p\log p-(1-p)\log(1-p) )\leq  n\log p \log (1-p).$ Thus by Theorem \ref{thr:mon}, there is an infinite sequence $\alpha\in\IS$ compatible with $S$, with $\K(\alpha)\lel \K(\mathcal{B}_p) + n\log p \log (1-p)+ \I(S;\ch)\lel \K(p) + n\log p \log (1-p)+ \I(S;\ch)$.

\end{exm}

\section{Games}
Function derandomization has applications to the cybernetic agent model, whose connection to Algorithmic Information Theory is studied extensively in \cite{Hutter05}.
In this section, we describe two simplified cybernetic agent models. For the first model, the agent $\p$ and environment $\q$ are defined as follows. The agent is a function $\p:(\N\times\N)^*\rightarrow\N$, where if $\p(w)=a$, $w\in (\N\times\N)^*$ is a list of the previous actions of the agent and the environment, and $a\in\N$ is the action to be performed. The environment is of the form $\q:(\N\times\N)^*\times\N\rightarrow \N\cup\{\mathbf{W}\}$, where if $\q(w,a)=b\in\N$, then $b$ is $\q$'s response to the agent's action $a$, given history $w$, and the game continues. If $\q$ responds $\mathbf{W}$ then the agents wins and the game halts. The agent can be randomized. The game can continue forever, given certain agents and environments. This is called a win/no-halt game.
\begin{thr}[\cite{Levin16,Epstein19}]
\label{thr:el}$ $\\
For finite $D\subset\FS$, $-\log\max_{x\in D}\m(x)\lel -\log\sum_{x\in D}\m(x)+\I(D;\ch)$.
\end{thr}
The following theorem is a game-theoretic interpretation of Lemma 6 in \cite{VereshchaginVi10}.
\begin{thr}
If $2^r$ deterministic agents of complexity $<k$ win against environment $\q$, then there is a deterministic agent $\p$ of complexity $\lel k-r+\I(\langle r,k,\q\rangle;\ch)$ that wins against $\q$.
\end{thr}
\begin{prf}
Given $\langle r,k,\q\rangle$, one can construct a finite set $D$ of encoded agents that win against $\q$ and $D$ contains at least $2^r$ agents of complexity $<k$. Furthermore $\sum_{x\in D}\m(x)\gem 2^r2^{-k}$, so using Theorem \ref{thr:el}, there is an agent $\p\in D$, where, using Lemma \ref{lmm:consH}, $\K(\p)\lel  k-r + \I(D;\ch)\lel  k-r + \I(\langle r,k,\q\rangle;\ch)$. \qed
\end{prf}

\begin{thr}
\label{thr:probwin}
If probabilistic agent $\p'$ wins against environment $\q$ with at least probability $p$, then there is a deterministic agent $\p$ of complexity $\lel \K(\p') -\log p + \I(\langle p, \p',\q\rangle;\ch)$ that wins against $\q$.
\end{thr}
\begin{prf}
Let $\mathcal{I}$ be a set of interactions between an arbitrary agent and the environment $\q$ such that each interaction ends in $\mathbf{W}$ and with probability $>p/2$, $\p'$ will act according to an interaction in $\mathcal{I}$. Thus $\K(\mathcal{I}|p,\p',\q)=O(1)$. $\p'$ can be encoded into a random function $F$, where the domain $(\N\times\N)^*$ of $\p'$ can be encoded into a single number $\N$. $\K(F|\p')=O(1)$. Similarly, $\mathcal{I}$ can be encoded into a set of samples $\mathfrak{C}$, where $\Pr[F(\mathfrak{C})]>p/2$ and $\K(\langle \mathfrak{C}\rangle|\langle\mathcal{I}\rangle)=O(1)$. Using Corollary \ref{cor:setsample}, there is a deterministic function $G:\N\rightarrow\N$, such that 
\begin{align}
\nonumber
\K(G)&\lel \K(F) -\log[F(\mathfrak{C})] + \I(\langle \mathfrak{C}\rangle;\ch) \\
\nonumber
&\lel \K(\p') -\log[F(\mathfrak{C})] + \I(\langle \mathfrak{C\rangle};\ch) \\
\nonumber
&\lel \K(\p') -\log p+ \I(\langle \mathfrak{C}\rangle;\ch) \\
\label{eq:gapplycon1}
&\lel \K(\p') -\log p+ \I(\langle \mathcal{I}\rangle;\ch) \\
\label{eq:gapplycon2}
&\lel \K(\p') -\log p+ \I(\langle p, \p',\q\rangle;\ch),
\end{align}
where Equations \ref{eq:gapplycon1} and \ref{eq:gapplycon2} are due to Lemma \ref{lmm:consH}. The deterministic function $G$ is an encoding of an agent, $\p$, proving the theorem.\qed
\end{prf}

\begin{exm}[Graph Navigation]
\label{exm:nav}
Let $G = (E,V)$ be a graph consisting of vertices $V$ and undirected edges $E$. By \cite{Lovasz96}, if $G$ is non-bipartite,  a random walk starting from any vertex will converge to a stationary distribution $\pi(v) = \mathrm{deg}(v)/2|E|$, for each $v \in V$. Let $t_G$ be the time it takes for any random walk starting anywhere to converge to the stationary distribution $\pi(v)$, for all $v \in V$, up to a factor of 2.

The win/no-halt game is as follows. The environment $\q$ consists of $(G,s,r)$. $G=(E,V)$ is a non-bipartite graph with undirected edges, $s \in V$ is the starting vertex, and $r \in V$ is the goal vertex. 

There are $t_G$ rounds and the agent starts at $s\in V$. At round 1, the environment gives the agent the degree $s\in V$, $\mathrm{Deg}(s)$. The agent picks an number between 1 and $\mathrm{Deg}(s)$ and sends it to $\q$. The agent moves along the edge the number is mapped to and is given the degree of the next vertex it is on. This process is repeated $t_G$ times. The agent wins if it is on $r\in V$ at the end of round $t_G$.

A probabilistic agent $\p$ is defined as selecting each edge with equal probability. After $t_G$ rounds, the probability that $p$ is on the goal $r$ is close to the stationary distribution $\pi$. More specifically the probability is $\gem 1/|E|$. Thus by Theorem \ref{thr:probwin}, there is a deterministic agent $\p'$ that can find $r$ in $t_G$ turns and has complexity $\K(\p')\lel \log |E| + \I((G,s,t);\ch)$. 

The game can even be generalized to have the environment choose each round's mapping of numbers to edges to be a function of the current vertex, round number, and the agent's past actions.
\end{exm}

The second game is modified such that the environment gives a nonnegative rational penalty term to the agent at each round. Furthermore the environment specifies an end to the game without specifying a winner or loser. This is called a penalty game.

\begin{cor}
\label{cor:pengame}
If given probabilistic agent $\p$, environment $\q$ halts with probability 1, and 
 $\p$ has expected penalty less than $n\in\N$, then there is a deterministic agent of complexity $\lel \K(\p) +\I(\langle \p,n,\q\rangle;\ch)$ that receives penalty $<2n$ against $\q$.
\end{cor} 
 \begin{prf}
We create a win/no-halt game from $\q$ where an agent wins if it gets a penalty less than $2n$. Thus $\p$ is a probabilistic agent that wins this new game with probability $>.5$. Theorem \ref{thr:probwin} then can be used to prove the corollary.\qed
\end{prf}$4$\\

The performance of the deterministic agent can increase at the cost of its complexity, by using the Markov inequality for different values.
\begin{exm}[Penalty Tests]
An example penalty game is as follows. The environment $\q$ plays a game for $N$ rounds, for some very large $N\in\N$, with each round starting with an action by $\q$. At round $i$, the environment gives, to the agent, a program to compute a probability $P_i$ over $\N$. The choice of $P_i$ can be a computable function of $i$ and the agent's previous turns. The agent responds with a number $a_i\in\N$. The environment gives the agent a penalty of size $T_i(a_i)$, where $T_i:\N\rightarrow\Q_{\geq 0}$ is a computable test, with $\sum_{a\in\N}P_i(a)T_i(a)<1$. After $N$ rounds, $\q$ halts.

A very successful probabilistic agent $\p$ can be defined. Its algorithm is simple. On receipt of a program to compute $P_i$, the agent randomly samples a number $\N$ according to $P_i$. At each round the expected penalty is $\sum_{a}P_i(a)T_i(a)<1$, so the expected penalty of $\p$ for the entire game is $<N$. Thus by Corollary \ref{cor:pengame}, there is a deterministic agent $\p'$ such that
\begin{enumerate}
\item $\p'$ receives a penalty of $<2N$,
\item  $\K(\p')\lel \I(\q;\ch)$.
\end{enumerate}

Let $\q$ be defined so that $P_i(a) = [a\leq 2^i]2^{-i}$ and $T_i = [a\leq 2^i]2^{i-\K(a|i)}$. Thus each $T_i$ is a randomness deficiency function. The probabilistic algorithm $\p$ will receive an expected penalty $<N$. However any deterministic agent $\p'$ that receives a penalty $<2N$ must be very complex, as it must select many numbers with low randomness deficiency. Thus, by the bounds above, $\I(\q;\ch)$ must be very high. This makes sense because $\q$ encodes $N$ randomness deficiency functions.
\end{exm}$ $\newpage

\section{Discussion}

Function derandomization can be applied to NP hard problems that admit randomized approximation algorithms. This can be seen in Examples \ref{exm:sat} and \ref{exm:cut} which deals with the {\sc Max-3Sat} and {\sc Max-Cut} problems, respectively. By using a randomized algorithm that achieves an approximation of the optimal result with positive probability, one can apply function derandomization to prove the existence of a deterministic algorithm that achieves the same score and has complexity $\lel\I(\langle \textrm{problem description}\rangle;\ch)$. If the algorithm requires a lot of information about the problem, then one can use games to show that a deterministic function will produce the approximate solution to the problem when fed the appropriate details about the problem. This interactive formulation can be seen in the graph navigation exercise of Example \ref{exm:nav}.

 In the proof of Theorem \ref{thr:dyn}, a relativization technique can be used to convert an $O(\K(x))$ error term to a $\K(x)$ error term, which allows the removal of quantifiers from the theorem statement. This technique can be performed by first relativizing inequalities to a shortest program that computes all the relevant parameters $\mu$, $\lambda$, and $n$. Then the next part is to reconfigure all terms that have the parameters as conditional information, in this case the deficiency of randomness $\D(\alpha|\mu)$. This technique was also used in \cite{EpsteinNO22,Epstein22}.
 

\end{document}